\documentclass[number]{elsarticle}
\usepackage{graphicx}% Include figure files
\usepackage{dcolumn}% Align table columns on decimal point
\usepackage{bm}% bold math

\begin{document}

%\preprint{APS/123-QED}

\title{Does H$_{2}$O improve the catalytic activity of Au$_{1-4}$/MgO towards CO oxidation?}
% Force line breaks with \\

\author[rvt]{Martin Amft\corref{cor1}}
\ead{martin.amft@fysik.uu.se}
\cortext[cor1]{Corresponding author}

\author[rvt]{Natalia V. Skorodumova}%

\address[rvt]{%
Department of Physics and Astronomy, Uppsala University, Box 516, SE-751 20 Uppsala, Sweden}%

\begin{abstract}
The present density functional theory study addresses the question whether the presence of H$_{2}$O influences the catalytic activity of small gold clusters, Au$_{1-4}$/MgO(100), towards the  oxidation of carbon monoxide.
To this end, we studied the (co-)adsorption of H$_{2}$O and CO/O$_{2}$ on these gold clusters.
The ground state structures in the presence of all three molecular species, that we found, are Au$_{1}$O$_{2}$/MgO and Au$_{2-4}$CO/MgO with H$_{2}$O adsorbed on the surface in the proximity of the  clusters-molecule complex.
In this configuration the catalytic activity of Au$_{1-4}$/MgO is indifferent to the presence of H$_{2}$O.
We also found that a stable, highly activated hydroperoxyl-hydroxyl complex, $\mathrm{O}_{2}\mathrm{H}\cdot\cdot\,\mathrm{OH}$, can be formed on Au$_{1,3}$/MgO.
For the catalytic active system Au$_{8}$/MgO, it has been predicted that this complex opens an alternative catalytic reaction pathway towards CO oxidation.
Our results suggest that this water mediated catalytic cycle is unlikely to occur on Au$_{1,3}$/MgO.
In the case of Au$_{1}$/MgO the cycle is interrupted by the dissociation of the remaining (OH)$_{2}$ complex after forming the first CO$_{2}$ molecule.
On Au$_{3}$/MgO the $\mathrm{O}_{2}\mathrm{H}\cdot\cdot\,\mathrm{OH}$ complex is likely to decompose  upon the impact of a CO molecule, since three of its bond dissociation energies are comparable to the reaction barrier of the CO to CO$_{2}$ oxidation.
\end{abstract}

%\pacs{68.47.Jn, 68.43.Fg, 68.35.Np, 82.65.+r}
% 68.47.Jn Clusters on oxide surfaces
% 68.43.Fg Adsorbate structure (binding sites, geometry
% 68.35.Np Adhesion
% 82.65.+r Surface and interface chemistry; heterogeneous catalysis at surfaces

%\keywords{Suggested keywords}%Use showkeys class option if keyword
                            %display desired


\maketitle

\section{\label{sec:intro}Introduction}
Gold, known and treasured for being chemically inert in bulk, received new attention after the discovery \cite{Haruta:1987p6162, Haruta:1997p6180} that supported nanometer-sized gold particles have a substantial catalytic activity towards low-temperature oxidation of carbon monoxide \cite{Arenz:2006p7702,Chretien:2007p6138, Coquet:2008p6144}.
Even sub-nanometer gold clusters, consisting of a few atoms only, showed an interestingly high activity towards CO oxidation, both in the gas phase and on metal-oxide supports \cite{Sanchez:1999p6314,Hakkinen:2001p6522,Socaciu:2003p7758,Yoon:2005p6472,Janssens:2007p9822}.
During the last decade much, predominantly theoretical, work has been carried out in this field, revealing for instance the importance of the cluster-substrate interaction \cite{Molina:2003p6216,Molina:2004p6219} and of the defects in the substrate \cite{Lopez:2004p10818,Barcaro:2008p11183}.
The influence of other chemical species on the catalytic activity of supported gold clusters has been studied, e.g. for chlorine, which hinders the catalytic activity  \cite{Broqvist:2004p9659}.

The effect of moisture on the catalytic activity of supported Au nanoclusters towards CO oxidation is less conclusive.
Both, enhanced activities of up to two orders of magnitude for Au/MgO \cite{Date:2004p10689} as well as inhibition of the reaction on Au/TiO$_{2}$ \cite{Gao:2010p10816} have been found in experiments.
Magnesium oxide is one of the most studied substrates used for gold cluster deposition because of its stability and simplicity to handle both in theory \cite{Skorodumova:2005p9330} and experiment \cite{Ferry:1998p11192}.
The catalytic activity of small gold clusters, Au$_{n}$ (n = 1,..., 20), towards CO oxidation has been observed \cite{Sanchez:1999p6314}.
Especially, Au$_{8}$/MgO gained much attention \cite{Yoon:2005p6472}.
So far, it is the only system of this type, where the influence of H$_{2}$O on its catalytic activity has been studied. 
In Ref. \cite{Bongiorno:2005p9732} it has been predicted that H$_{2}$O can open an alternative reaction pathway for the oxidation of carbon monoxide, hence promoting the activity of Au$_{8}$/MgO. 

Recently, the co-adsorption of CO and O$_{2}$ on Au$_{1-4}$/MgO(100) has been systematically  studied \cite{Amft:2010p11084}.
In that work the absence of catalytic activity towards CO oxidation of Au$_{1,2}$/MgO and the small activity of Au$_{3,4}$/MgO, observed experimentally in \cite{Sanchez:1999p6314}, could be explained by theory.

Here we address the question whether the presence of H$_{2}$O in these model systems influences the adsorption behavior of CO and O$_{2}$ on Au$_{1-4}$/MgO(100) and if a water promoted catalytic activity, as suggested in Ref. \cite{Bongiorno:2005p9732}, can be found even for the smallest gold clusters on a regular MgO terrace.
We concentrate our affords on Au$_{1-4}$/MgO, because of their restricted number of isomers, which allows us to reliably determine the ground state structures.
These sub-nanometer gold clusters are immobile on a regular MgO terrace at low temperatures due to their size-dependent diffusion barriers E$_{\mathrm{diff}}$ of 0.19 to 0.62 eV \cite{Barcaro:2007p10886} and adsorption energies E$_{\mathrm{ads}}$ of -0.89 to -1.75 eV \cite{Amft:2010p11084}.
Our model system is appropriate under the assumption of a low defect concentration on the MgO surface.

The adsorption and diffusion of a wide range of molecular species on regular MgO terraces has extensively been studied by theory.
The reported adsorption energies are -0.08 eV for CO \cite{Snyder:2000p11194} and -0.04 eV \cite{Geneste:2005p11191} for O$_{2}$, respectively.
Since CO and O$_{2}$ bind significantly stronger on Au$_{1-4}$/MgO than on bare MgO \cite{Amft:2010p11084}, we concentrate on the (co-)adsorption of these molecular species from the gas phase.
The H$_{2}$O molecule has been found to bind stronger to MgO and being less mobile, i.e. E$_{\mathrm{ads}}$ = -0.37 eV and E$_{\mathrm{diff}}$ = 0.13 eV \cite{Odelius:1999p10806,Alfe:2007p11193}.
Its adsorption on Au$_{1-4}$/MgO has not been studied so far. 

The paper is organized as follows.
After summarizing the computational details in section \ref{sec:comp}, we will, as a first step, present our results of the adsorption of H$_{2}$O on Au$_{1-4}$/MgO, section \ref{sec:waterads}.
This allows us, in a second step, to investigate the possible co-adsorption scenarios of CO or O$_{2}$ with H$_{2}$O on Au$_{1-4}$/MgO, section \ref{sec:coads}.
In section \ref{sec:catal} we address the question, whether a water promoted CO oxidation cycle, as  proposed in the case of Au$_{8}$/MgO,\cite{Bongiorno:2005p9732} is possible even for the smallest gold clusters.
Finally, section \ref{sec:summ} summarizes our results and concludes this work.

\section{Computational details}
\label{sec:comp}
The scalar-relativistic  \textit{ab-initio} DFT calculations were performed using the projector augmented wave (PAW) \cite{Blochl:1994p10844,Kresse:1999p10843} method as implemented in \textsc{vasp} \cite{Kresse:1996p6093,Kresse:1996p6092}.
The exchange-correlation interaction was treated in the generalized gradient approximation (GGA) in the parameterization of Perdew, Burke, and Ernzerhof (PBE) \cite{PERDEW:1996p6520}.
A cut-off energy of 600 eV was used and a Gaussian smearing of $\sigma$ = 0.02 eV  for the occupation of the electronic levels.
Spin-polarization was taken into account for all calculations.
The MgO(100) surface was  modeled in a super-cell approach with a two monolayer thick $3\times3$ MgO slab. 
The unit cell was constructed using the equilibrium lattice parameter of MgO (4.235\,\AA) obtained in the corresponding bulk calculations.
The repeated slabs were separated from each other by 27~\AA\, of vacuum.
A Monkhorst-Pack $\Gamma$-centered $2\times2\times1$ k-point mesh was used for the structural relaxations.
The relaxation cycle was stopped when the Hellmann-Feynman forces had become smaller than $5 \cdot 10^{-3}$ eV/{\AA}. 
The charges were calculated by means of the Bader analysis \cite{Tang:2009p9327}.
For an exothermic adsorption of A on adsorbent S the adsorption energy, $E_{\mathrm{ads}} = E_{0}\mathrm{[A/S]} - E_{0}\mathrm{[A]} - E_{0}\mathrm{[S]}$, is negative.

\section{Adsorption of H$_{2}$O on A\lowercase{u}$_{1-4}$/M\lowercase{g}O}
\label{sec:waterads}
The gold clusters Au$_{1-4}$ stand on the regular MgO(100) terrace and form bonds with oxygen atoms O$_{s}$, which donate 0.3 e$^{-}$ per Au-O$_{s}$ bond to the clusters \cite{Molina:2004p6219,Frondelius:2007p6471,Amft:2010p11084}.

\begin{figure}[tbh] 
\centering
\includegraphics[width=0.6\linewidth]{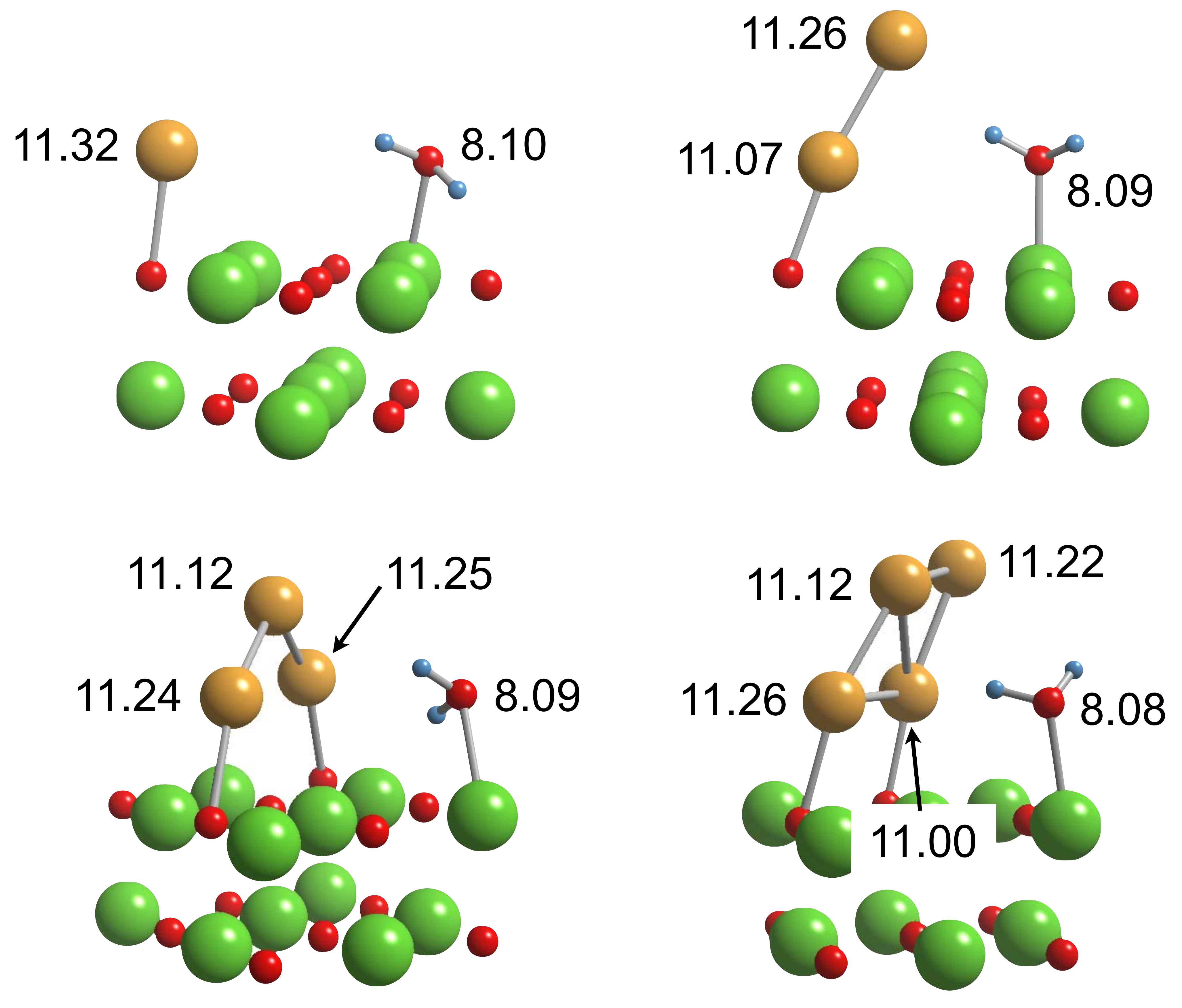}
\caption{(Color online) Ground state geometries of Au$_{1-4}$/MgO(100) with H$_{2}$O adsorbed on the surface in the proximity of the cluster. Note that H$_{2}$O prefers to bind to the next-nearest surface Mg. Also shown are the Bader charges of the gold and water-oxygen atoms (in e$^{-}$). The Bader charge of a neutral gold atom is 11 and that of the oxygen atom in H$_{2}$O is 8. The different atom species are illustrated as yellow (Au), red (O), green (Mg), and blue (H) balls, respectively.}
\label{fig:figure1}
\end{figure}

\begin{figure}[tbh] 
\centering
\includegraphics[width=0.6\linewidth]{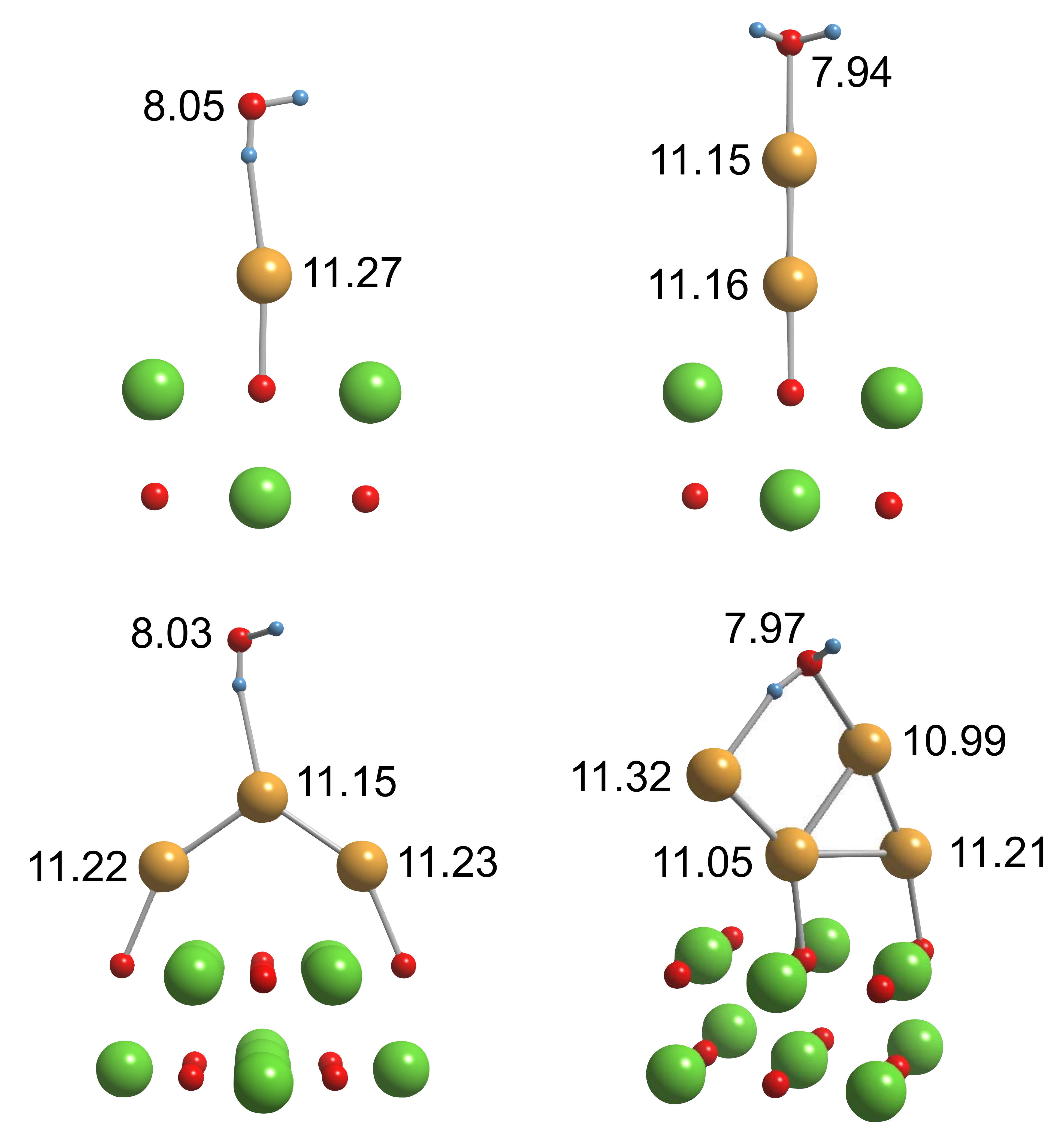}
\caption{(Color online) Ground state geometries of Au$_{1-4}$/MgO(100) with H$_{2}$O adsorbed  on top of the clusters. Note the different binding mechanisms depending on the odd-even number of gold atoms in the cluster. Also shown are the Bader charges of the gold and water-oxygen atoms (in e$^{-}$), cf. Fig. \ref{fig:figure1}. The different atom species are illustrated as yellow (Au), red (O), green (Mg), and blue (H) balls, respectively.}
\label{fig:figure2}
\end{figure}

\begin{figure}[hbt]
\centering
\includegraphics[width=0.8\linewidth]{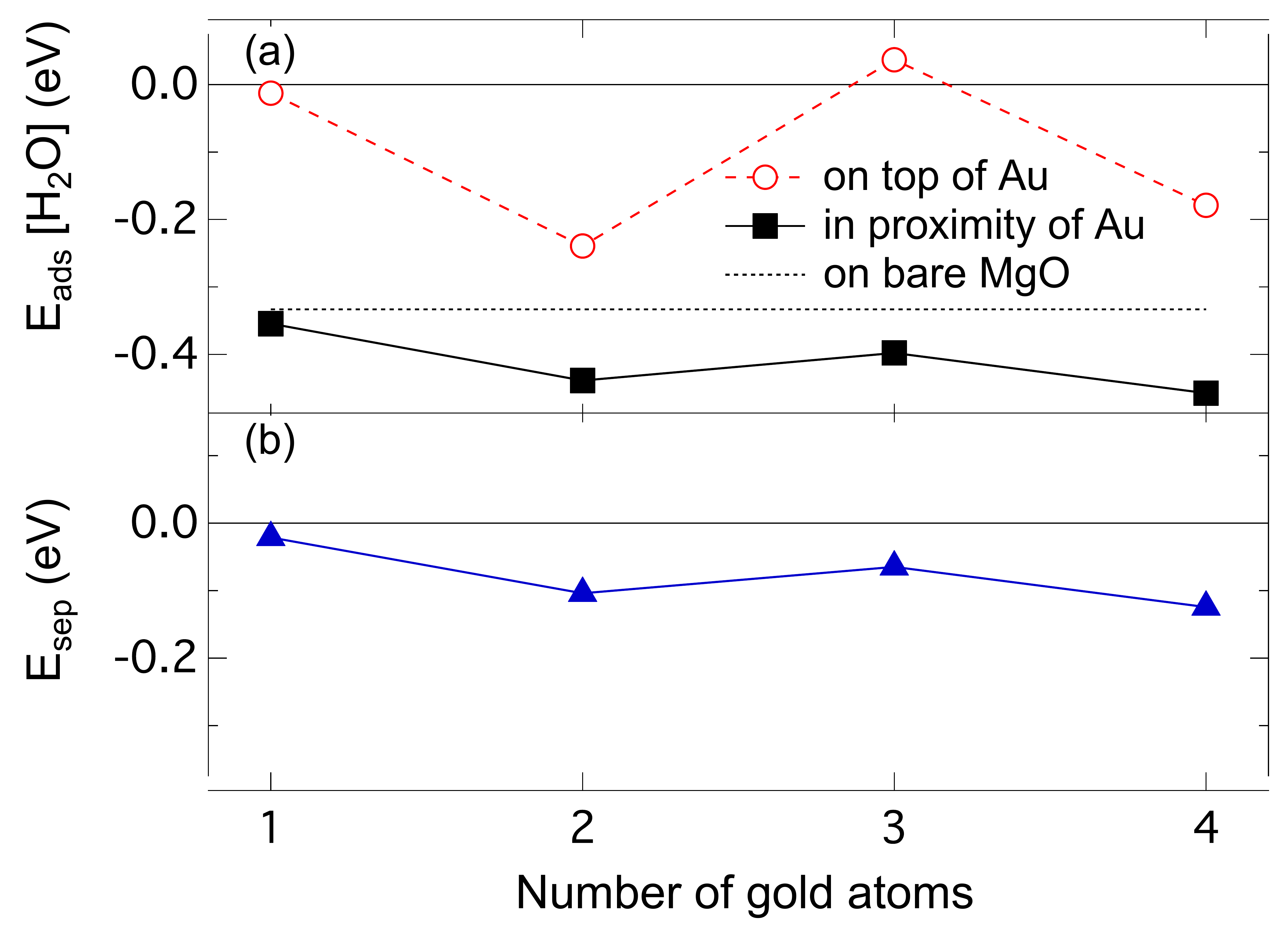}  
\caption{(Color online) Top panel (a): Adsorption energies of H$_{2}$O in the proximity (solid squares) and on top of Au$_{1-4}$/MgO (open circles). The dashed horizontal line shows E$_{\mathrm{ads}}$ of H$_{2}$O on bare MgO. Lower panel (b): the separation energy E$_{\mathrm{sep}}$ =  E$_{0}$[Au$_{n}$/MgO with H$_{2}$O in proximity] - E$_{0}$[Au$_{n}$/MgO] - E$_{0}$[H$_{2}$O/MgO] (solid triangles). Note that these points coincide with the solid squares in panel (a), i.e. E$_{\mathrm{ads}}$ [Au$_{1-4}$] is not affected by the proximity of H$_{2}$O.}
\label{fig:figure3}
\end{figure}

Starting from various initial configurations, we identified two possible sets of adsorption sites for H$_{2}$O. 
First, on the MgO surface in the proximity of Au$_{1-4}$, Fig. \ref{fig:figure1}, and, second, on top of Au$_{1-4}$/MgO, Fig. \ref{fig:figure2}.
The adsorption energies for both cases are shown in Fig. \ref{fig:figure3} (a) together with E$_{\mathrm{ads}}$ on bare MgO.

We found that the binding mechanisms of H$_{2}$O in the proximity of Au$_{1-4}$ and on bare MgO are rather similar: in both cases the water-oxygen atom binds to a surface magnesium Mg \cite{Langel:1995p10733,Almeida:1998p10807,Odelius:1999p10806}. 
It receives a small addition charge of up to 0.1 e$^{-}$ from the surface oxygens surrounding this Mg.
The charge donations from the surface oxygens to the gold clusters and the water molecule are virtually unaffected by their mutual proximity.
Also the structural changes in the Au clusters are rather small, i.e. bond lengths change by less than 1.5\%. 

In the proximity of Au$_{1-4}$/MgO two effects influence the adsorption energy of H$_{2}$O: first, the Coulomb repulsion between the negatively charged cluster and molecule and, second, the orientation of the water molecule's electrical dipole towards the clusters.
At the binding site on top of the next-nearest Mg from the gold cluster H$_{2}$O gains most energy from the combination of these two effects.

The energy cost of separating H$_{2}$O from the Au cluster, E$_{\mathrm{sep}}$ =  E$_{0}$[Au$_{n}$/MgO with H$_{2}$O in proximity] - E$_{0}$[Au$_{n}$/MgO] - E$_{0}$[H$_{2}$O/MgO] shown in Fig. \ref{fig:figure3} (b), is identical to the difference E$_{\mathrm{ads}}^{\mathrm{prox}}$ - E$_{\mathrm{ads}}^{\mathrm{bare}}$ of the H$_{2}$O adsorption energies close and further away from the cluster.
Therefore the adsorption energy of the gold clusters is not affected by the proximity of the water molecule and the energy is gained from the orientation of the electrical dipole, only.

The other (meta-)stable adsorption sites for H$_{2}$O are on top of the gold clusters, see Fig. \ref{fig:figure2} and Fig. \ref{fig:figure3}(a) for the adsorption energies.
The adsorption on top of the clusters shows pronounced cluster-size effects: depending on the parity of the number of Au atoms in the cluster H$_{2}$O forms an Au-H (odd) or an Au-O bond (even), see Fig. \ref{fig:figure2}.
The oscillations of properties due to the changing parity of the number of Au atoms were previously observed for low dimensional gold structures \cite{Skorodumova:2005p10022,Grigoriev:2006p10023}.
The different binding mechanisms for the top sites are readily explained by the electronic structures of Au$_{1-4}$/MgO: gold clusters with an odd number of atoms have an unpaired 6$s$ electron that forms a weak bond with hydrogen's 1$s$ electron. 
On the other hand the 5$d$-states are the highest occupied states of the clusters consisting of an even number of atoms, forming a stronger bond with the 2$p$ electrons of oxygen.

These different odd-even bonding mechanisms are, for instance, also reflected in the oscillation of the gold-surface oxygen bond length d$_{\mathrm{Au-O}_{s}}$ and the charge redistributions to the water molecule.
For Au$_{1,3}$ d$_{\mathrm{Au-O}_{s}}$ decreases by 2.5\%, whereas it increases by 2\% for Au$_{2,4}$. 
The most remarkable water-induced structural change is the breaking of the Au-Au bond between the two top gold atoms of Au$_{4}$ with H$_{2}$O in the top position.
Adsorbed on top of Au$_{1,3}$ up to 0.05 e$^{-}$ are redistributed to the water molecule, while the molecule loses up to 0.06 e$^{-}$ when adsorbed on  Au$_{2,4}$, see the Bader charges in Fig. \ref{fig:figure2}.

\section{Co-adsorption of H$_{2}$O and CO/O$_{2}$ on A\lowercase{u}$_{1-4}$/M\lowercase{g}O}
\label{sec:coads}

\begin{figure}[hbt]
\centering
\includegraphics[width=0.8\linewidth]{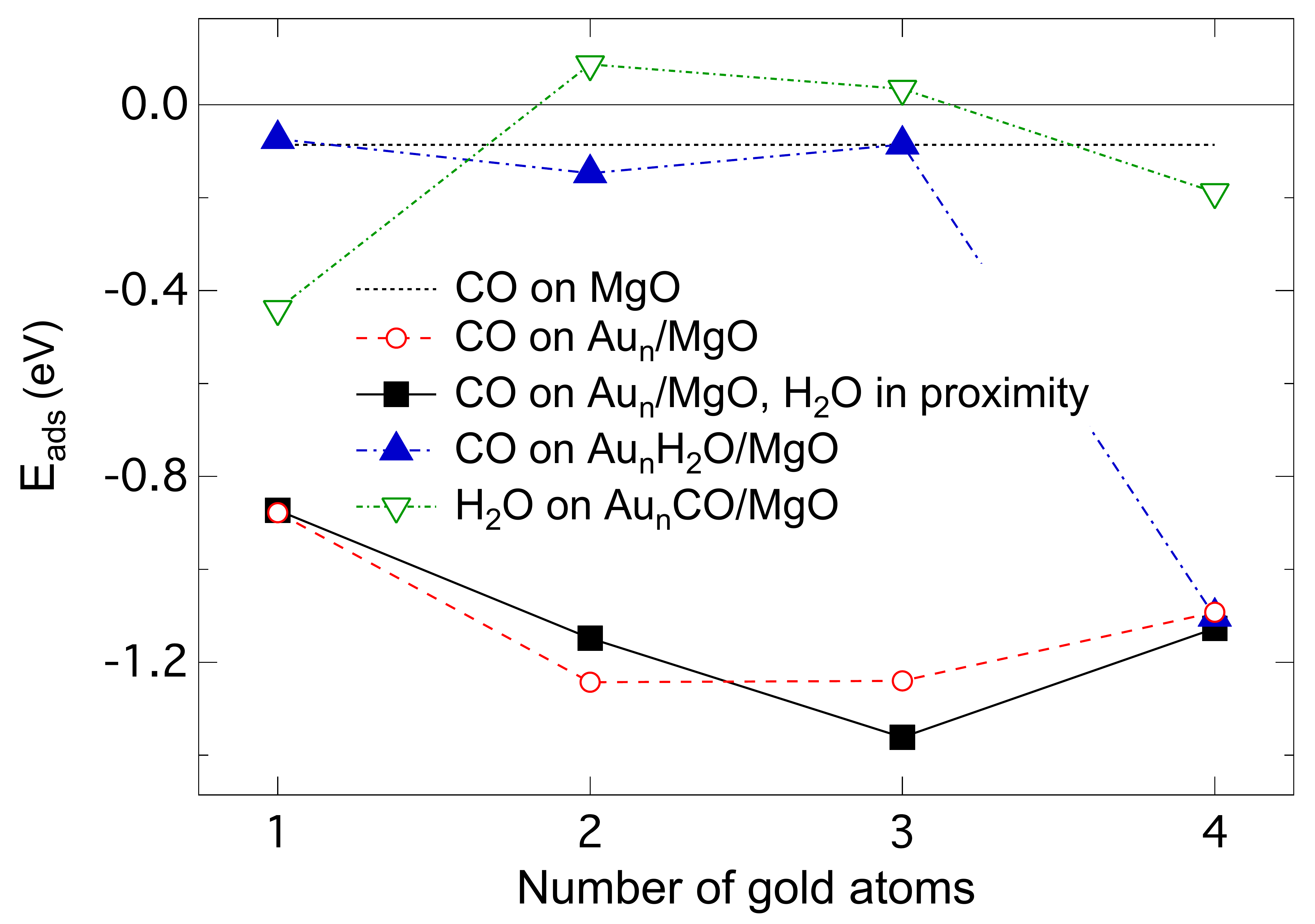}  
\caption{(Color online) Adsorption energies of CO on bare MgO (dotted horizontal line) and on Au$_{1-4}$/MgO (open circles). In the presence of a water molecule different co-adsorption scenarios are possible: H$_{2}$O pre-adsorbed in the proximity of Au$_{1-4}$/MgO (solid squares), H$_{2}$O pre-adsorbed on top of Au$_{n}$/MgO (solid triangles), and H$_{2}$O adsorbed on top of Au$_{1-4}$CO/MgO (open triangles).}
\label{fig:figure4}
\end{figure}

\begin{figure}[tbh] 
\centering
\includegraphics[width=0.8\linewidth]{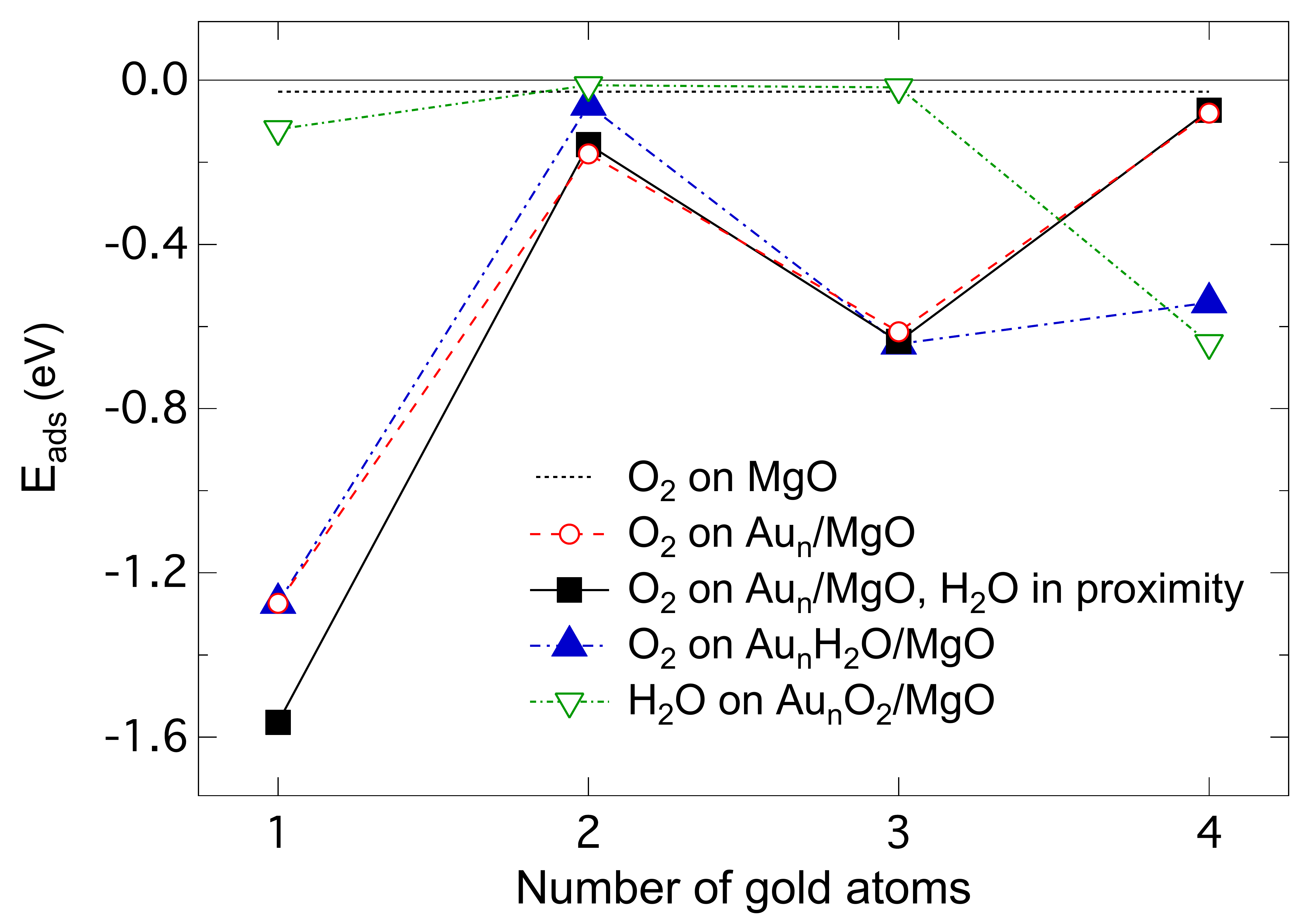} 
\caption{(Color online) Adsorption energies of O$_{2}$ on bare MgO (dotted horizontal line) and on Au$_{1-4}$/MgO (open circles). In the presence of a water molecule different co-adsorption scenarios are possible: H$_{2}$O pre-adsorbed in the proximity of Au$_{1-4}$/MgO (solid squares), H$_{2}$O pre-adsorbed on top of Au$_{n}$/MgO (solid triangles), and H$_{2}$O adsorbed on top of Au$_{1-4}$O$_{2}$/MgO (open triangles).}
\label{fig:figure5} 
\end{figure}

Recently the (co-)adsorption of CO and O$_{2}$ on Au$_{1-4}$/MgO(100) has been systematically studied \cite{Amft:2010p11084}.
Here, based on the analysis of the individual adsorption energies of H$_{2}$O, CO, and O$_{2}$, we can determine the cluster-molecule ground state structures in a mixed atmosphere of all three molecular species.
In Figures \ref{fig:figure4} and \ref{fig:figure5} the adsorption energies of CO and O$_{2}$ on Au$_{1-4}$/MgO with and without of H$_{2}$O are compiled.

As mentioned above, CO and O$_{2}$ adsorb weakly on a regular MgO terrace, most favorably on top of Mg, with E$_{\mathrm{ads}}$ = -0.09 and -0.03 eV, respectively (see dotted horizontal lines in Figs. \ref{fig:figure4} and \ref{fig:figure5}).
On top of the gold clusters Au$_{1-4}$/MgO, on the other hand, CO and O$_{2}$ bind stronger (open circles in Figs. \ref{fig:figure4} and \ref{fig:figure5}) \cite{Amft:2010p11084}.
The general trend is, both with and without H$_{2}$O, that CO binds significantly stronger on top of Au$_{2-4}$/MgO than oxygen does.

\begin{figure*}[hbt]
\centering
\includegraphics[width=0.95\linewidth]{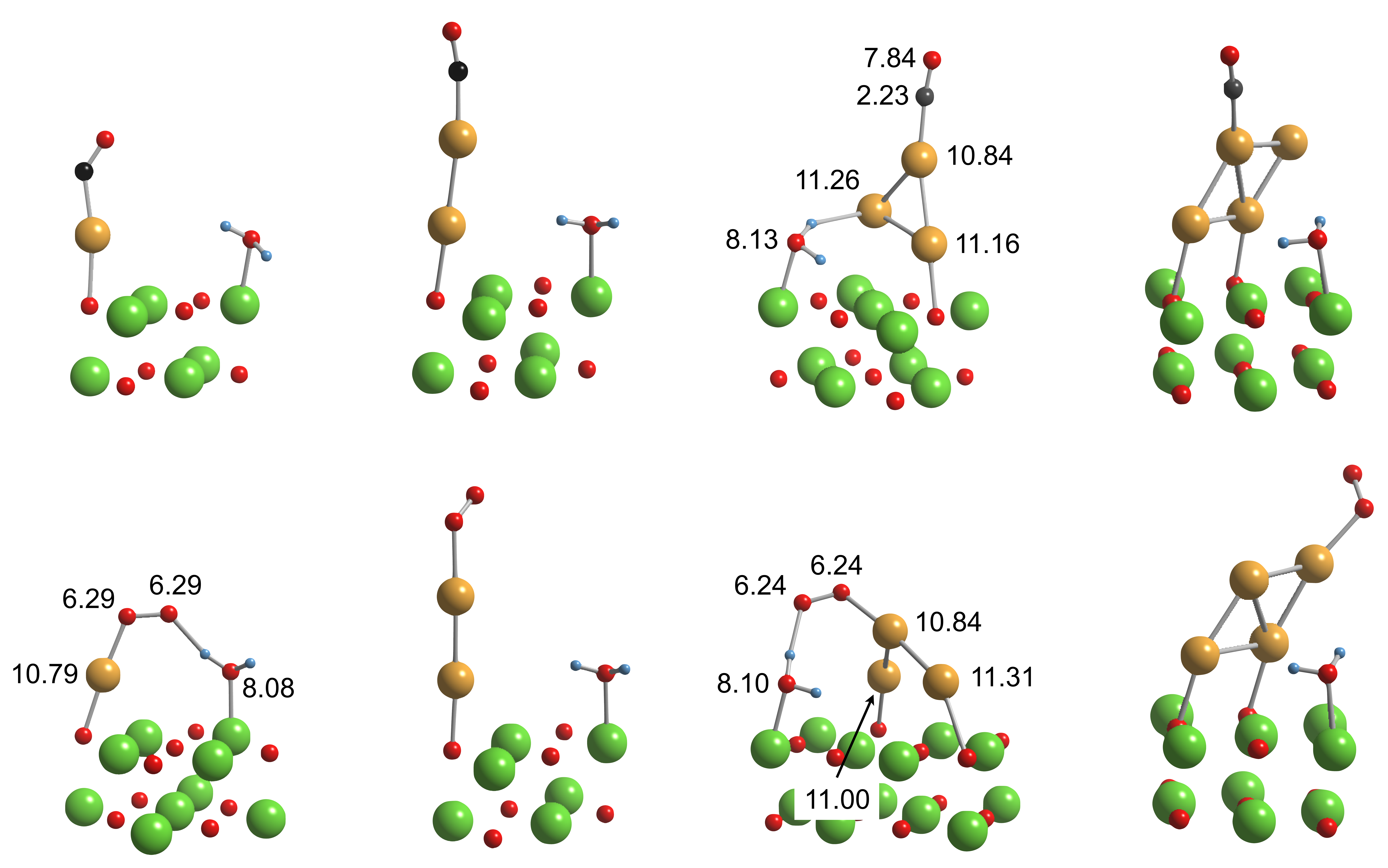}  
\caption{(Color online) Ground state structures of CO/O$_{2}$ and H$_{2}$O co-adsorbed on Au$_{1-4}$/MgO. In the three cases were new bonds are formed also the Bader charges are shown (in e$^{-}$), cf. Figs. \ref{fig:figure1} and \ref{fig:figure2}. In the gas phase carbon monoxide has Bader charges of 2.25 (C) and 7.75 (O) and the oxygen molecule of 6 per O-atom. The different atom species are illustrated as yellow (Au), red (O), black (C), green (Mg), and blue (H) balls, respectively.}
\label{fig:figure6}
\end{figure*}

As we showed in the previous section, the most favorable adsorption sites for H$_{2}$O are in the proximity of the gold clusters, on top of the next-nearest Mg atom.
The solid squares in Figs. \ref{fig:figure4} and \ref{fig:figure5} show the E$_{\mathrm{ads}}$[CO] and E$_{\mathrm{ads}}$[O$_{2}$] on top of Au$_{1-4}$/MgO with H$_{2}$O in the proximity.
Some structures are energetically indifferent to the proximity of H$_{2}$O, e.g. Au$_{1,4}$CO and Au$_{2-4}$O$_{2}$.
In other cases CO or O$_{2}$ bind stronger to the gold, e.g. Au$_{3}$CO and Au$_{1}$O$_{2}$.
Only in the case of the dimer CO binds weaker with H$_{2}$O in the proximity.
Figures \ref{fig:figure6} illustrates the ground state structures of CO/O$_{2}$ adsorbed on Au$_{1-4}$/MgO with H$_{2}$O in the proximity.
The water molecule leads only in three of the systems to notable structural changes.
First, the breaking of a Au-O$_{s}$ bond in favor of a Au-H bond for Au$_{3}$CO/MgO, which is accompanied by charge redistribution within the cluster.
Second, Au$_{1,3}$O$_{2}$/MgO bend down towards H$_{2}$O and form a H-O bond.
In Au$_{3}$O$_{2}$/MgO it also leads to a charge redistribution within the cluster, i.e. one of the Au atoms close to the surface loses its additional charge.

Although the adsorption sites of H$_{2}$O on top of the gold structures are energetically less favorable than on the surface, Fig. \ref{fig:figure3} (a), we study the possibility for CO and O$_{2}$ to co-adsorb on top of Au$_{1-4}$H$_{2}$O/MgO (solid triangles in Figs. \ref{fig:figure4} and \ref{fig:figure5}).
Carbon monoxide adsorbs on Au$_{1,3}$H$_{2}$O/MgO as weakly as on the bare MgO surface.
On the even-numbered clusters CO can bind stronger than on MgO (Au$_{2}$) or essentially as strong as in the absence of H$_{2}$O (Au$_{4}$).
The tetramer has two adsorption sites, the two gold atoms at the top. 
If CO and H$_{2}$O co-adsorb on one of these sites each, they barely influence each other, as it can be seen, for instance, from their unaltered adsorption energies.

The oxygen molecule can only weakly co-adsorb on Au$_{2}$H$_{2}$O/MgO, which does not lead to any structural changes.
In the other three cases O$_{2}$ forms, to different degrees, new molecular complexes with the pre-adsorbed H$_{2}$O.
Especially on the odd-numbered clusters hydroperoxyl-hydroxyl complexes, i.e. O$_{2}$H$\cdot\cdot$OH, are formed on top of the gold clusters, Fig. \ref{fig:figure7}, by moving a hydrogen atom towards the oxygen molecule.
Although the co-adsorption of O$_{2}$ and H$_{2}$O on the two adsorption sites of Au$_{4}$ has a synergetic effect on their E$_{\mathrm{ads}}$, see Fig. \ref{fig:figure5}, a hydroperoxyl-hydroxyl complex, as on Au$_{1,3}$, is not formed.
The hydrogen atom remains in the H$_{2}$O molecule with a bond length stretched by 3\% and also the intramolecular bond in O$_{2}$ is only stretched by 6\%.

To complete this study of  co-adsorption, we also address the question of H$_{2}$O co-adsorbing on pre-adsorbed CO/O$_{2}$ (open triangles in Figs. \ref{fig:figure4} and \ref{fig:figure5}).
From the adsorption energies obtained for these scenarios one sees that H$_{2}$O has a meta-stable local minimum on top of Au$_{2,3}$CO/MgO and binds weakly on top of Au$_{2,3}$O$_{2}$/MgO, in the latter case without a significant additional activation of O$_{2}$.
On Au$_{1}$CO/O$_{2}$ the water molecule can bind, although without a significant additional activation of the molecules.
Co-adsorbed on the second adsorption site on top of Au$_{4}$ we found the highest E$_{\mathrm{ads}}$[ H$_{2}$O] with -0.64 eV.

In conclusion, we predict the formation of the following ground state structures under a mixed atmosphere: Au$_{1}$O$_{2}$/MgO and Au$_{2-4}$CO/MgO with a water molecule binding on the surface in the proximity of the cluster, cf. Figs. \ref{fig:figure3} to \ref{fig:figure5}.
Except for the dimer the adsorption energies of CO and O$_{2}$ are enhanced by the proximity of a water molecule.
Since the preferential gold/CO and gold/O$_{2}$ structures are similar to those found for the water free case \cite{Amft:2010p11084}, we conclude that also the blocking of Au$_{1,2}$/MgO by O$_{2}$ and CO, respectively, as well as the CO/O$_{2}$ co-adsorption on Au$_{3,4}$/MgO are essentially unaltered.
Therefore also the catalytic activities of these systems are indifferent to the proximity of H$_{2}$O.

\section{Is a H$_{2}$O mediated CO oxidation reaction possible on Au$_{1-4}$/MgO?}
\label{sec:catal}

In Ref. \cite{Bongiorno:2005p9732} a H$_{2}$O mediated reaction pathway towards CO oxidation has been suggested for Au$_{8}$/MgO.
The proposed catalytic cycle consists of the following steps: 
\begin{equation}
\label{Eq1}
\mathrm{Au}_{n}\mathrm{/MgO} + \mathrm{H}_{2}\mathrm{O}  + \mathrm{O}_{2} \rightarrow \mathrm{Au}_{n}\mathrm{O}_{2}\mathrm{H}\cdot\cdot\,\mathrm{OH/MgO}
\end{equation}
\begin{equation}
\label{Eq2}
\mathrm{Au}_{n}\mathrm{O}_{2}\mathrm{H}\cdot\cdot\,\mathrm{OH/MgO} + \mathrm{CO} \rightarrow \mathrm{Au}_{n}\mathrm{(OH)}_{2}\mathrm{/MgO} + \mathrm{CO}_{2}
\end{equation}
\begin{equation}
\label{Eq3}
\mathrm{Au}_{n}\mathrm{(OH)}_{2}\mathrm{/MgO} + \mathrm{CO} \rightarrow \mathrm{Au}_{n}\mathrm{/MgO} + \mathrm{H}_{2}\mathrm{O}  + \mathrm{CO}_{2}.
\end{equation}

It essentially depends on the formation of a highly activated complex, consisting of a hydroperoxyl and a hydroxyl group, i.e. $\mathrm{O}_{2}\mathrm{H}\cdot\cdot\,\mathrm{OH}$, on top of the gold cluster.
Although we found that these complexes are stable on top of Au$_{1,3}$/MgO, see E$_{\mathrm{ads}}$ in Fig. \ref{fig:figure5}, their formation probability is limited by the fact that H$_{2}$O has only a meta-stable adsorption site on Au$_{1,3}$/MgO, Fig. \ref{fig:figure3} (a).

\begin{figure}[hbt]
\centering
\includegraphics[width=0.6\linewidth]{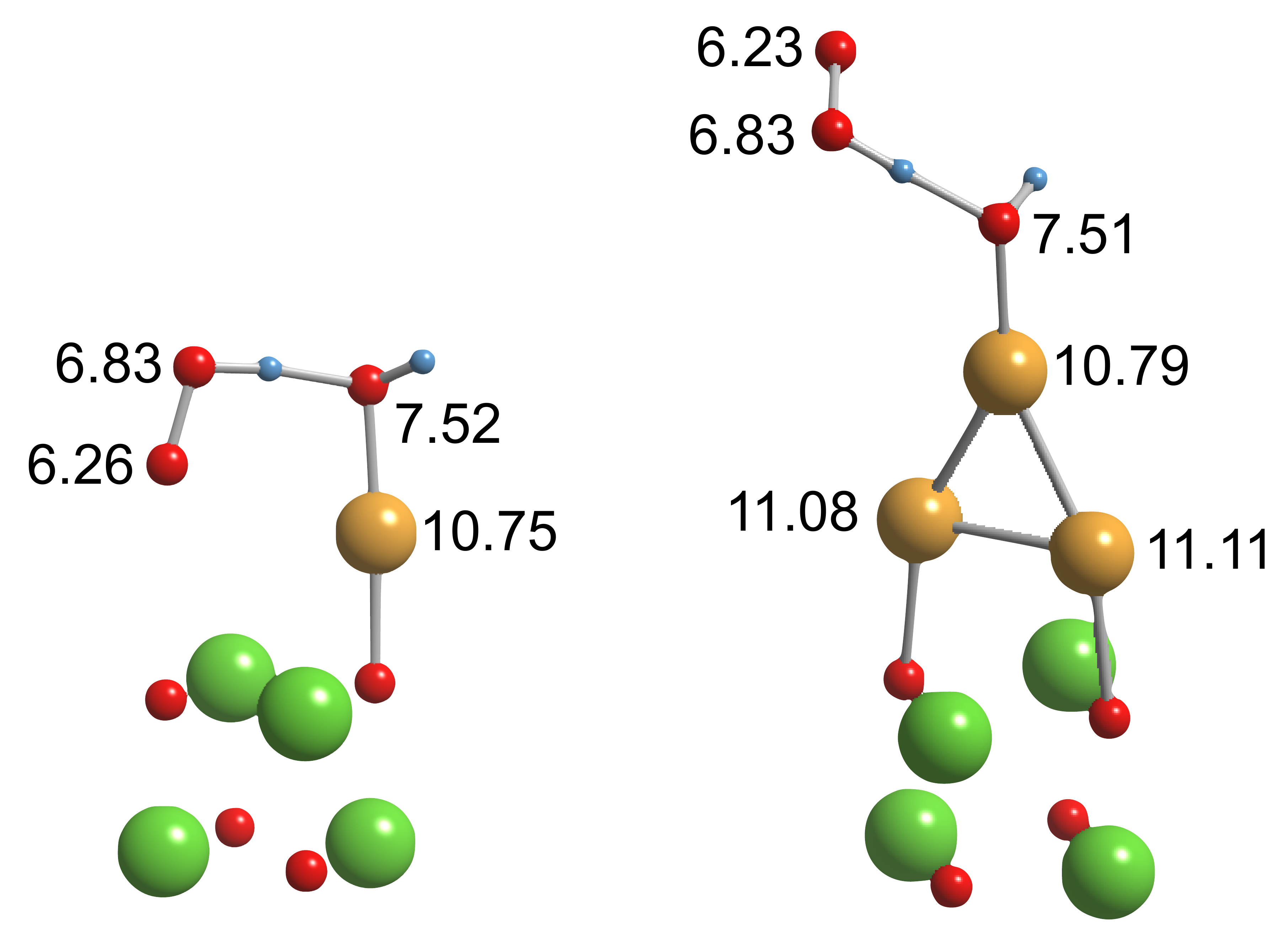}  
\caption{(Color online) On top of Au$_{1,3}$/MgO highly activated complexes, consisting of a hydroperoxyl and a hydroxyl group, i.e. $\mathrm{O}_{2}\mathrm{H}\cdot\cdot\,\mathrm{OH}$, can be formed from an oxygen molecule and a pre-adsorbed water molecule, cf. Fig. \ref{fig:figure5}. The Bader charges are shown as well (in e$^{-}$). The different atom species are illustrated as yellow (Au), red (O), green (Mg), and blue (H) balls, respectively.}
\label{fig:figure7}
\end{figure}

Figure \ref{fig:figure7} shows the ground state structure of these two complexes and the calculated Bader charges.
One hydrogen atom is moved from the pre-adsorbed water molecule towards the oxygen molecule, which redistributes its electron towards O$_{2}$.
The so formed hydroperoxyl group has a highly activated O-O bond, i.e. the bond is stretched by 9\% to 1.34 \AA.

The remaining hydroxyl group receives an additional charge of 0.5 e$^{-}$ from the gold clusters,  i.e. a Bader charge of 7.5 e$^{-}$ on OH instead of 7 e$^{-}$, Fig. \ref{fig:figure7}.

\begin{figure}[hbt]
\centering
\includegraphics[width=0.95\linewidth]{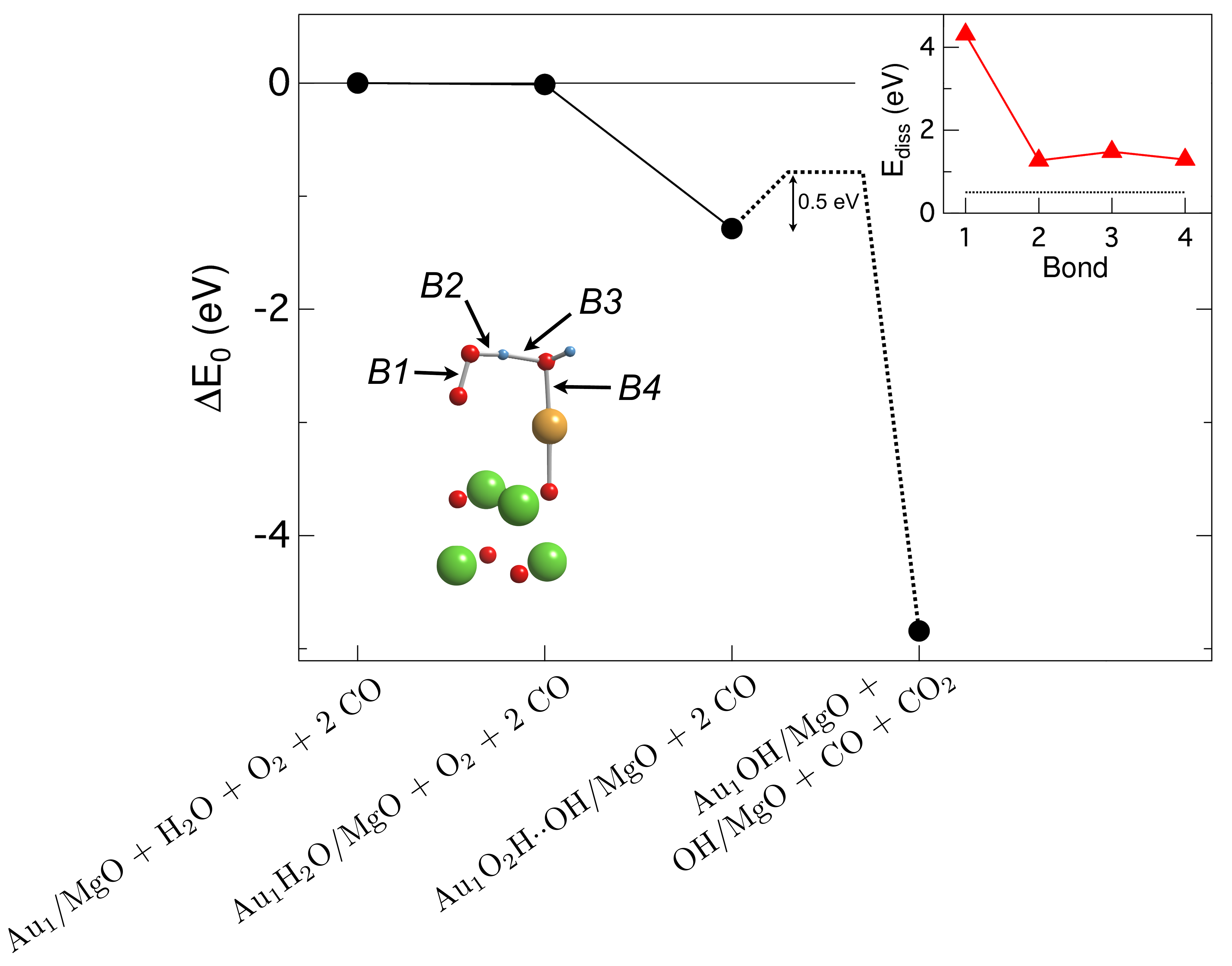}  
\caption{(Color online) Change in the total energy during the successive adsorption of H$_{2}$O, O$_{2}$, and CO on Au$_{1}$/MgO. After the formation of a CO$_{2}$ the remaining complex break apart into two OH groups, one on the adatom and one directly on the surface nearby. The upper-right inset shows the bond dissociation energies of B1 - B4 (solid triangles) compared to the activation barrier (dotted line) of the $\mathrm{O}_{2}\mathrm{H}\cdot\cdot\,\mathrm{OH}$ complex, formed in step 3 and illustrated in the lower-left corner.}
\label{fig:figure8}
\end{figure}

\begin{figure}[hbt]
\centering
\includegraphics[width=0.95\linewidth]{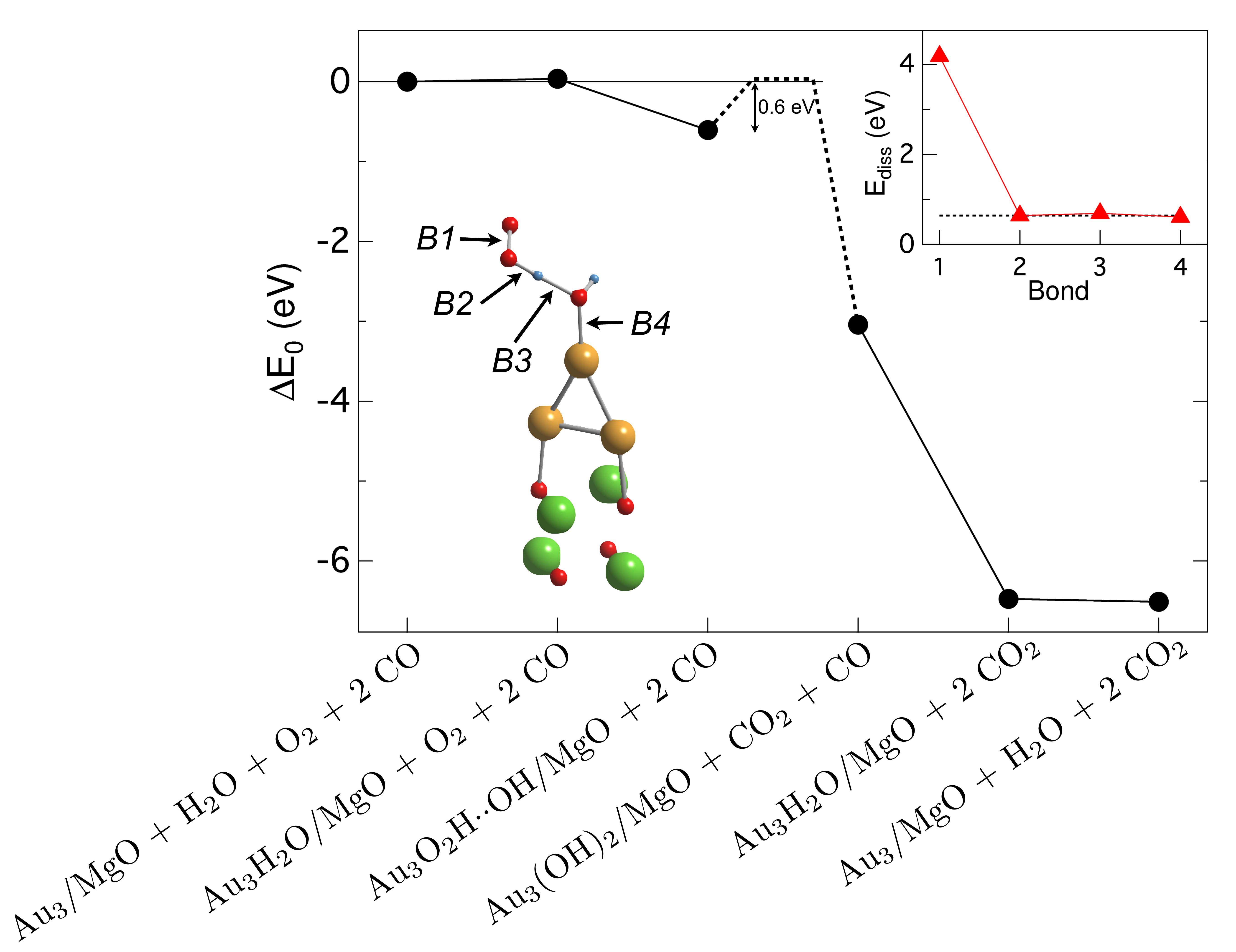}  
\caption{(Color online) Change in the total energy during the successive adsorption of H$_{2}$O, O$_{2}$, and CO on Au$_{3}$/MgO to model the full catalytic cycle. The inset in the upper-right corner  shows the bond dissociation energies of B1 - B4 (solid triangles) compared to the activation barrier (dotted line) of the $\mathrm{O}_{2}\mathrm{H}\cdot\cdot\,\mathrm{OH}$ complex, formed in step 3 and illustrated in the lower-left corner.}
\label{fig:figure9}
\end{figure}

Naturally the question arises, if these complexes open a reaction pathway towards CO oxidation as in the case of Au$_{8}$/MgO?
To answer this, we calculated the total energies of the system during the reaction steps of the catalytic cycle Eqs (\ref{Eq1}) to (\ref{Eq3}), see Figs. \ref{fig:figure8} and \ref{fig:figure9}.

For Au$_{1,3}$/MgO, where the $\mathrm{O}_{2}\mathrm{H}\cdot\cdot\,\mathrm{OH}$ can be formed, we find it unlikely that a full catalytic cycle can be established.
The formation of the first CO$_{2}$ molecule from a CO molecule and an oxygen atom from the hydroperoxyl group involves an activation barrier.
We obtained activation energies for this reaction of 0.5 eV for Au$_{1}$ and 0.6 eV for Au$_{3}$  from a restricted molecular dynamics simulation.
The kinetic energy of the CO molecule has to exceed these activation energies.
Thus, a necessary condition for the oxidation reaction to happen is that the strength of the bonds B2 - B4 in the molecular complex, shown in the insets of Figs \ref{fig:figure8} and \ref{fig:figure9}, exceeds the kinetic energy of CO. 
This might occur for Au$_{1}$/MgO, since the remaining bonds B2 - B4 in the $\mathrm{O}_{2}\mathrm{H}\cdot\cdot\,\mathrm{OH}$ complex have bond dissociation energies of at least 1.27 eV.
However, the remaining complex (OH)$_{2}$ on Au$_{1}$/MgO undergoes a barrier free dissociation into two hydroxyl groups.
One of which is adsorbed on the adatom and the other one nearby on the MgO surface.
This dissociation is possible since bond B3 is weakened to -0.6 eV once the oxygen atom is removed from the hydroperoxyl group.
On MgO, on the other hand, the adsorption energy of an OH group is -1.19 eV.
Both OH are inert towards a reaction with the second CO molecule, therefore interrupting the catalytic cycle.

In the case of Au$_{3}$/MgO the situation is different.
First of all the adsorption energy of the $\mathrm{O}_{2}\mathrm{H}\cdot\cdot\,\mathrm{OH}$ complex is more than 50 \% smaller than that on the adatom.
At the same time the activation barrier for the formation of CO$_{2}$ is higher, making this step even slightly endothermic, cf. Fig. \ref{fig:figure9}.
This reaction becomes even more problematic, if one considers the bond dissociation energies of B2 - B4, cf. insets in Fig. \ref{fig:figure9}, which are very close to the activation barrier. 
Hence it is quite likely that one of the bonds B2 - B4 breaks upon the impact of a CO molecule with a kinetic energy, necessary to overcome the reaction barrier.
Still, if we assume for a moment that the bonds B2 - B4 will hold and the first CO$_{2}$ molecule is formed, than the remaining steps in the proposed catalytic cycle are exothermic and barrier free, see Fig. \ref{fig:figure9}. 

In conclusion, the dissociation of (OH)$_{2}$ in the case of Au$_{1}$ and the probable breaking of one of the bonds B2 - B4 in the molecular complex on Au$_{3}$ makes the proposed water mediated catalytic cycle unlikely to occur on the small gold cluster considered here.

\section{\label{sec:summ}Summary and conclusions}
By means of density functional theory we studied the adsorption of H$_{2}$O on small gold clusters, Au$_{1-4}$, supported by a regular MgO terrace.
We found that even these small gold clusters attract water molecules to their proximity.
Two sets of stable adsorption sites exist: first, the energetically more favorable site on MgO in the proximity of the gold clusters, and, second, on top of the clusters.
The latter showed a pronounced odd-even oscillation in the binding mechanism between H$_{2}$O and gold, depending on the number of gold atoms in the cluster.

We systematically studied the co-adsorption of H$_{2}$O with CO and O$_{2}$ on Au$_{1-4}$/MgO and found that it is  in general energetically more favorable to have CO/O$_{2}$ adsorbed on top of the cluster with H$_{2}$O in the proximity.
Although we found the adsorption energies of CO and O$_{2}$ to be enhanced by the proximity of H$_{2}$O, the known small catalytic activities of Au$_{1-4}$/MgO towards CO oxidation remain unaltered in this configuration.

On Au$_{1,3}$/MgO the oxygen and water molecules can form a stable, highly activated $\mathrm{O}_{2}\mathrm{H}\cdot\cdot\,\mathrm{OH}$ complex.
In contrast to the previously published findings for Au$_{8}$/MgO, these complexes are unlikely to open a catalytic reaction pathway towards CO oxidation.
The reaction barrier for the first CO oxidation is 0.6 eV on the gold trimer, which is comparable to the bond dissociation energies in three of the five bonds in the $\mathrm{O}_{2}\mathrm{H}\cdot\cdot\,\mathrm{OH}$ complex.
Hence, the molecular complex is likely to decompose as a result of the CO impact.

Although the bonds are stronger in the $\mathrm{O}_{2}\mathrm{H}\cdot\cdot\,\mathrm{OH}$ complex on the adatom, the intermediate (OH)$_{2}$ complex, formed by oxidizing one CO molecule, dissociates barrier free into two hydroxyl groups, one on the adatom and one on the surface nearby.
Since these OH-groups are found to be inert towards CO, the catalytic cycle is interrupted.

We conclude that H$_{2}$O can promote the co-adsorption of CO and O$_{2}$ on Au$_{1-4}$/MgO, but a stable, H$_{2}$O mediated, catalytic cycle towards CO oxidation is unlikely to occur on these clusters when supported by a defect free MgO terrace.
In real-world samples surface defects, like steps and vacancies, might be responsible for catalytic activities of sub-nanometer sized gold clusters.
Further investigations into this field are desirable.

\section*{Acknowledgments}
This research was supported by the Swedish Energy Agency (Energimyndigheten), the Swedish Research Council (VR), Research and Innovation for Sustainable Growth (VINNOVA), and the Swedish National Infrastructure for Computing (SNIC).

\bibliographystyle{elsarticle-num}
\bibliography{references}% Produces the bibliography via BibTeX.

\end{document}